\documentclass[namedreferences]{solarphysics}
\usepackage[optionalrh]{spr-sola-addons} 
\usepackage{graphicx}        
\usepackage{color}           
\usepackage{url}             

\begin{document}

\begin{article}

\begin{opening}

\title{Validity of the relations between the synodic and sidereal rotation velocities of the Sun}

\author{I.~\surname{Skoki\'{c}}$^{1}$\sep
             R.~\surname{Braj\v sa}$^{2}$\sep
             D.~\surname{Ro\v sa}$^{3}$\sep
             D.~\surname{Hr\v zina}$^{3}$\sep
             H.~\surname{W\"ohl}$^{4}$
       }
\runningauthor{Skoki\'{c} et al.}
\runningtitle{Synodic and sidereal rotation velocity of the Sun}

   \institute{$^{1}$ Cybrotech Ltd, Zagreb, Croatia
                     email: \url{ivica.skokic@gmail.com}\\ 
                   $^{2}$ Hvar Observatory, Faculty of Geodesy, University of Zagreb, Croatia
                     email: \url{romanb@geof.hr} \\
                   $^{3}$ Zagreb Astronomical Observatory, Zagreb, Croatia
                     email: \url{drosa@zvjezdarnica.hr} email: \url{dhrzina@zvjezdarnica.hr} \\
                   $^{4}$ Kiepenheuer-Institut f\"ur Sonnenphysik, Freiburg, Germany
                     email: \url{hw@kis.uni-freiburg.de}
             }

\begin{abstract}
Existing methods for conversion between synodic and sidereal rotation velocities of the Sun are tested for validity using state of the art ephemeris data. It is found that some of them are in good agreement with ephemeris calculations, while the other ones show a discrepancy of almost $0.01^{\circ}$ day$^{-1}$. The discrepancy is attributed to a missing factor and a new corrected relation is given.
\end{abstract}
\keywords{Rotation}
\end{opening}

\section{Introduction}
     \label{S-Introduction} 

Among important problems in the determination of the solar rotation is the measurement precision of various observing methods \cite{Schroter1985,Beck2000}. One example of that is the question of the influence of tracer's height on the measured rotation rate ({\it e.g.}, \opencite{Vrsnak1999}). Another important issue is the correct application of the synodic to sidereal rotation velocity transformation \cite{Graf1974,Rosa1995,Wittmann1996,Brajsa2002}. 

Due to the motion of Earth around the Sun, the observed (synodic, apparent) rotation period of the Sun is different from the real (sidereal, true) period. The relationship between the two is even more complicated by the seasonal effect resulting from the elliptic motion of the Earth and the inclination of the solar rotation axis to the ecliptic. If not taken into account, this effect can introduce a systematic error in studies of the solar rotation.

There have been several methods presented over the last few decades for the calculation of sidereal from synodic rotation period \cite{Graf1974,Rosa1995,Wittmann1996}. Hereafter, we discuss the validity of those methods by comparing them with best available ephemeris data, because these methods are frequently used by researchers and synodic to sidereal transformation is an essential step in solar rotation studies ({\it e.g.}, \opencite{Ye1998}; \opencite{Godoli1998}; \opencite{Brajsa1999}; \opencite{Zaatri2006}; \opencite{Komm2009}; \opencite{Heristchi2009}; \opencite{Wohl2010}; \opencite{Hiremath2013}). It is found that relations by \inlinecite{Graf1974} and \inlinecite{Rosa1995} are in general agreement with ephemeris calculations, while that of \inlinecite{Wittmann1996} shows a discrepancy of almost $0.01^{\circ}$ day$^{-1}$, as will be shown in this paper.  
 
\section{Relation between the synodic and sidereal rotation velocity} 
      \label{S-general}      

Sidereal solar rotation velocity, $\omega_{sid}$, was usually calculated by adding the mean orbital angular velocity of the Earth, $\overline{\omega}_{Earth}$, to the measured synodic velocity $\omega_{syn}$:
\begin{equation} \label{e:basic}
  \omega_{sid}=\omega_{syn}+\overline{\omega}_{Earth}.
\end{equation}

For the $\overline{\omega}_{Earth}$ a value of $0.9856^{\circ}$ day$^{-1}$  is usually used. \inlinecite{Graf1974} pointed out that this approach leads to periodic errors up to $0.15^{\circ}$ day$^{-1}$ because it does not include the effects of ellipticity of the Earth's orbit and inclination of solar rotation axis to the ecliptic, and suggested a new relation which takes into account these effects. His method was later improved by \inlinecite{Rosa1995} (with further corrections in \opencite{Brajsa2002}):
\begin{equation} \label{e:rosa}
  \omega_{sid}=\omega_{syn}+\left \{ \arctan [\cos i \tan(\lambda_{0}^{\prime}-\Omega^{\prime})]-\arctan [\cos i \tan(\lambda_0-\Omega)] \right \}.
\end{equation}
where $\lambda_0$ is the apparent longitude of the Sun referred to the true equinox of the date, $\Omega$ is the longitude of the ascending node of the solar equator on the ecliptic and $i$ is the inclination of the solar equator on the ecliptic. Primed variables denote values one day after the non-primed values. \inlinecite{Rosa1995} also gave approximate equations to calculate $\lambda_0$ and $\Omega$ without the need for printed ephemerides.

\begin{figure}
  \centerline{\includegraphics[width=0.6\textwidth,clip=]{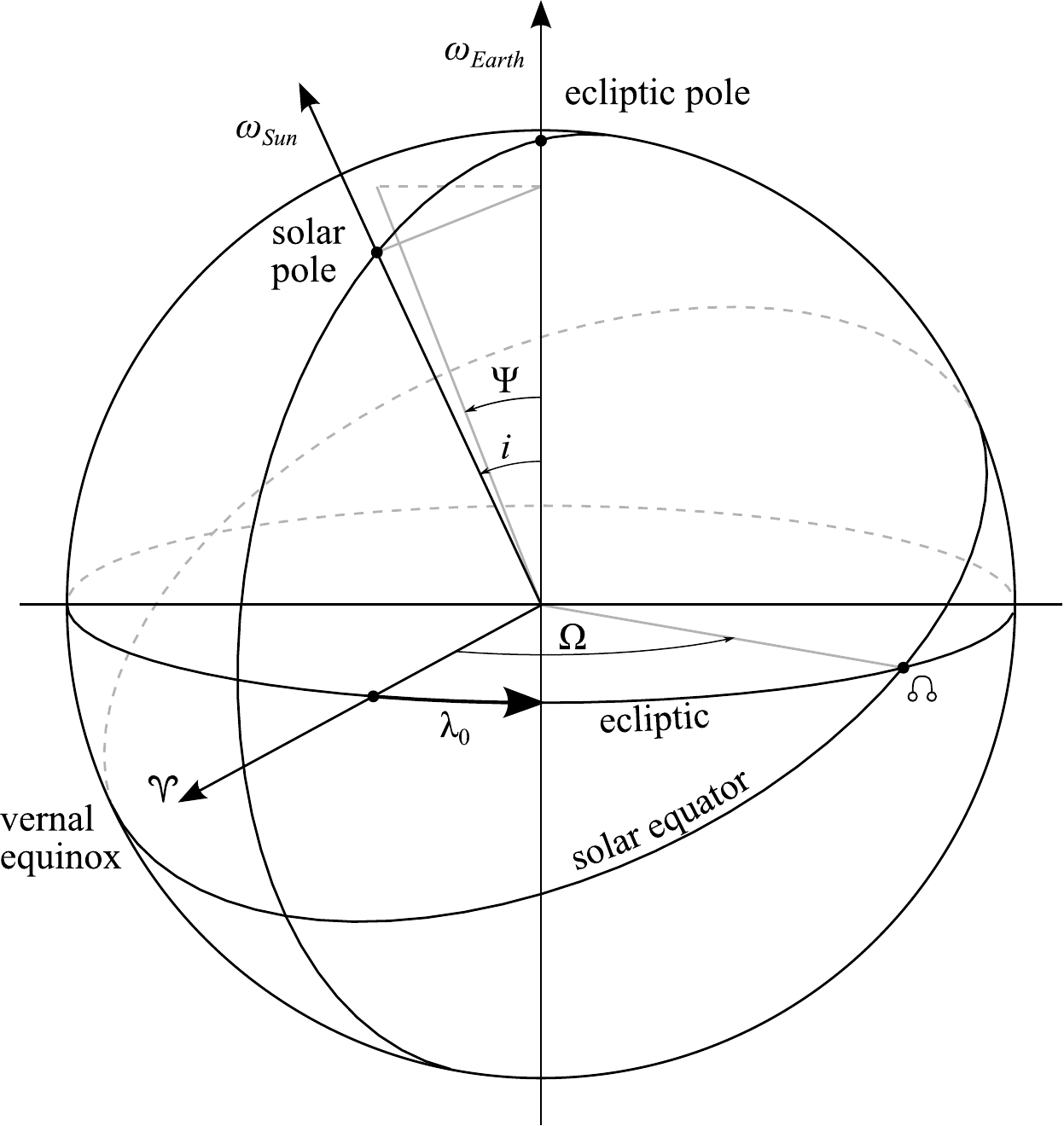}}
  \caption{Explanation of the angles used in the paper. The Earth is positioned towards the reader.} 
  \label{fig:sphere}
\end{figure}

\begin{figure}
  \centerline{\includegraphics[width=0.9\textwidth,clip=]{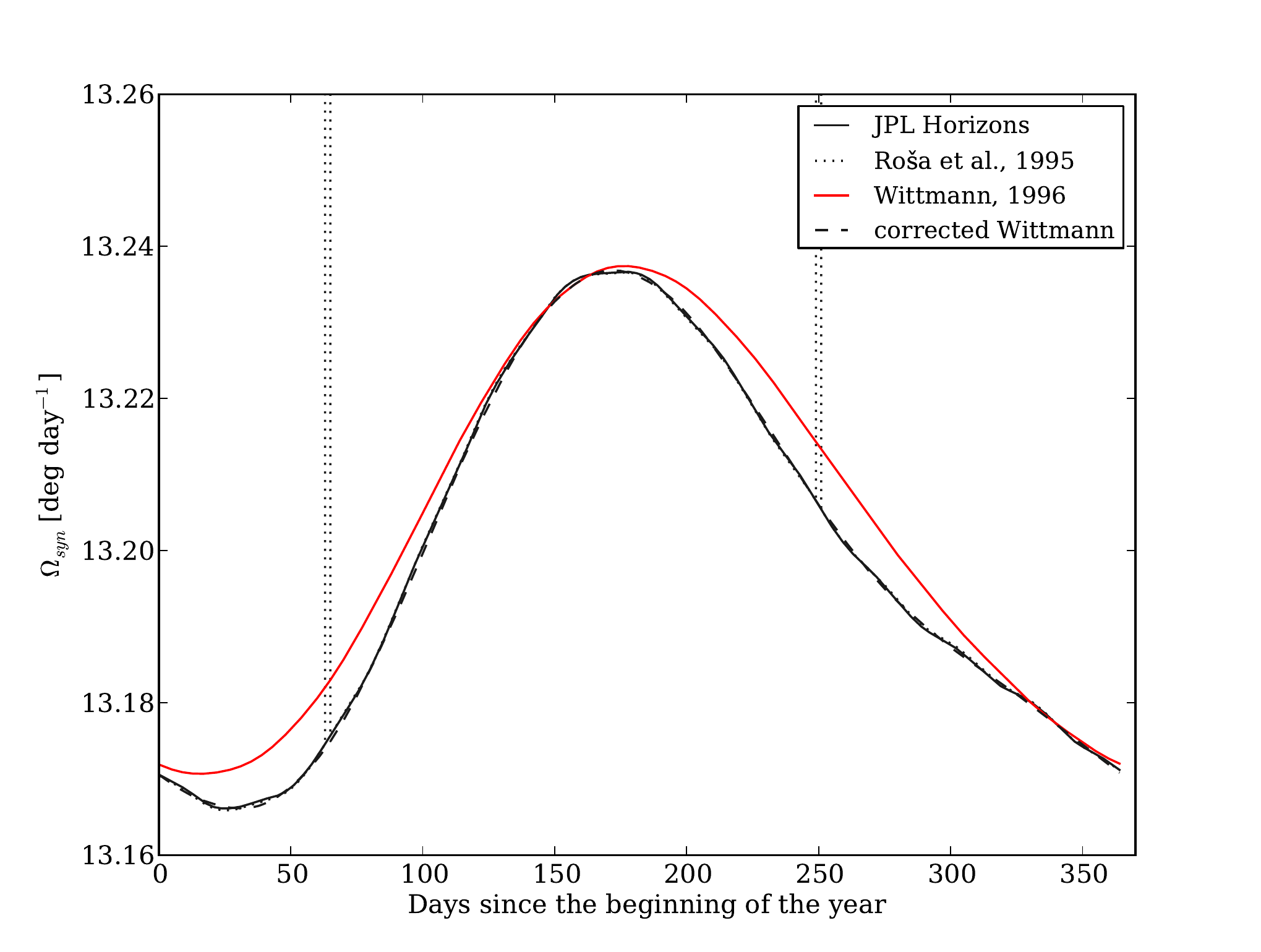}}
  \caption{Synodic velocity as a function of time as calculated by different methods for a fixed sidereal velocity of $14.1844^{\circ}$ day$^{-1}$. Method by \protect\inlinecite{Wittmann1996} gives slightly different results from the others which show practically the same behavior. Also shown are discontinuities in method of \protect\inlinecite{Rosa1995} resulting when arctangent functions are not handled carefully.} 
  \label{fig:syn-vel}
\end{figure}

Equation (\ref{e:rosa}) has two disadvantages. First, the attribution of the correct quadrant should be taken into account when calculating arctangent functions, otherwise the discontinuities could emerge and lead to wrong results (see Figure \ref{fig:syn-vel}). This problem was discussed in \inlinecite{Brajsa2002}. Second, calculation of $\lambda_0$ and $\Omega$ for two consecutive days is needed. A simpler method which avoids this was proposed by \inlinecite{Wittmann1996}:
\begin{equation} \label{e:wittmann}
  \omega_{sid}=\omega_{syn}+\omega_{Earth}\cos \Psi,
\end{equation}
where $\omega_{Earth}$ is the (instantaneous) orbital angular velocity of the Earth and $\Psi$ is the angle between the pole of the ecliptic and the solar rotation axis orthographically projected onto the solar disk (see Figure \ref{fig:sphere}). This angle can be calculated from \inlinecite{Very1897}:
\begin{equation} \label{e:psi}
  \tan \Psi=\tan i \cos (\lambda_0-\Omega),
\end{equation}
while $\omega_{Earth}$ can be found with sufficient accuracy using the Kepler's second law ({\it e.g.}, \opencite{Goldstein1983}, Equation (3-53)):
\begin{equation} \label{e:omega}
  \omega_{Earth}=\frac{\overline{\omega}_{Earth}}{r^2},
\end{equation}
where $r$ is the distance between the Earth and the Sun, expressed in AU. We also note a typing error in \inlinecite{Wittmann1996} where on several occasions a value of $0.9865^{\circ}$ day$^{-1}$ instead the correct value $0.9856^{\circ}$ day$^{-1}$ was printed for $\overline{\omega}_{Earth}$. 

\begin{figure}
  \centerline{\includegraphics[width=0.9\textwidth,clip=]{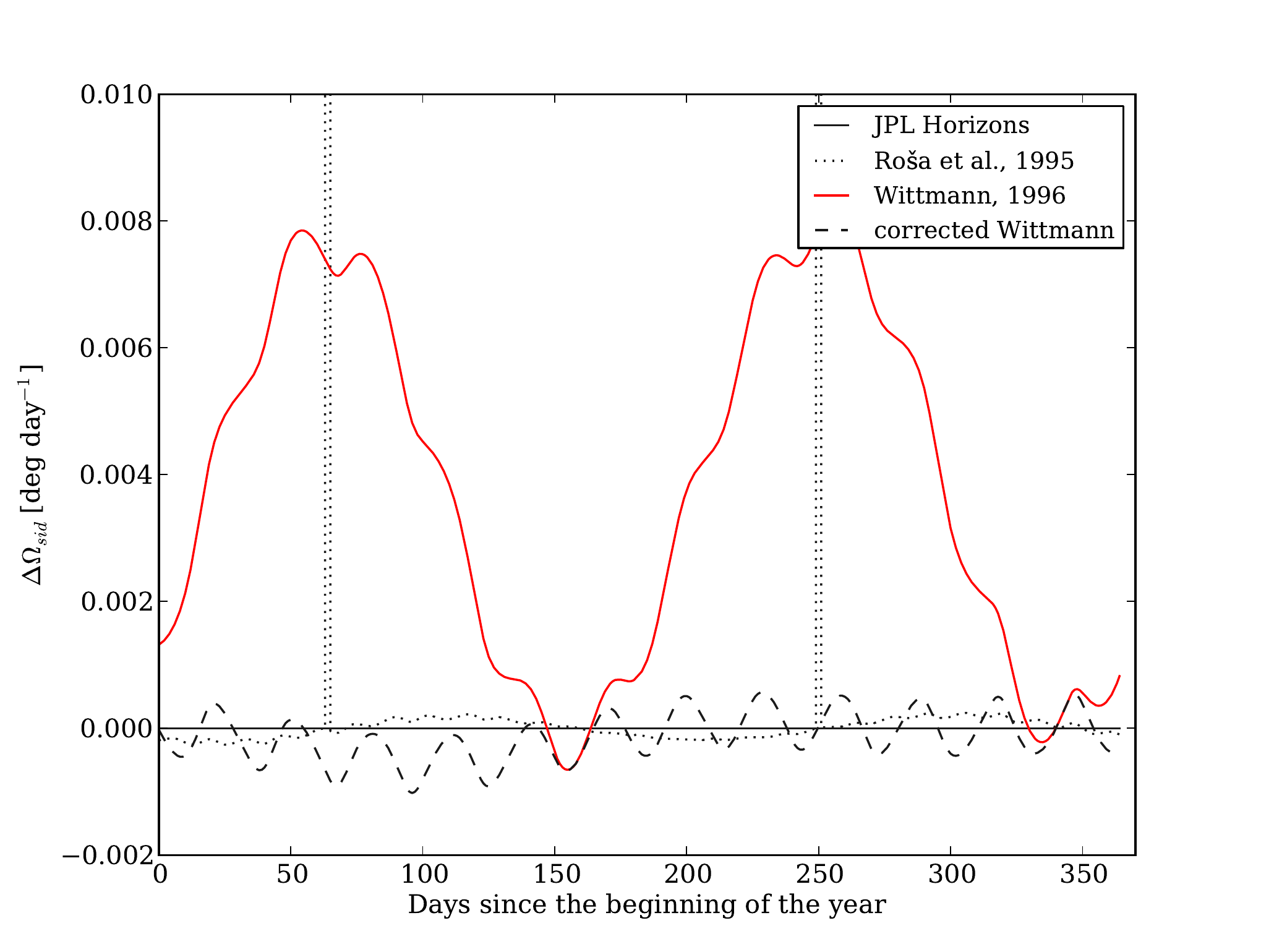}}
  \caption{Residual errors as a function of time for different methods of calculation of sidereal velocity.} 
 \label{fig:syn-res}
\end{figure}

\section{Calculations and results}
To investigate the validity of the given relations, we used ephemeris data from JPL Horizons system\footnote{\url{http://ssd.jpl.nasa.gov/?horizons}} \cite{JPL1996}, which we used to calculate heliographic longitude of the Earth, $L_0$, and instantaneous sidereal-synodic correction by using the following relation:
\begin{equation}  \label{e:corr}
  \omega_{sid}-\omega_{syn}=\omega_C+\frac{dL_0}{dt},
\end{equation}
where $\omega_{C}=14.1844^{\circ}$ day$^{-1}$ is the Carrington rotation rate. This correction was applied to the fixed sidereal rotation velocity, $\omega_{sid}$, for which we also used the Carrington rate with a period of 25.38 days, to get the reference synodic velocity. Also, the synodic velocities were calculated by Equations (\ref{e:rosa}) and (\ref{e:wittmann}), and the results were compared (Figure \ref{fig:syn-vel}). It can be seen that method by \inlinecite{Rosa1995} agrees well with the JPL ephemeris data, while that of Wittmann shows a small discrepancy. 

A careful check of Equation (\ref{e:wittmann}) shows that a factor is missing that re-projects orthographically the projected orbital angular velocity of the Earth back to the solar rotation axis. If this is done, one gets the following relation:
\begin{equation} \label{e:this}
  \omega_{sid}=\omega_{syn}+\omega_{Earth}\frac{\cos^2 \Psi}{\cos i}.
\end{equation}

As can be seen from Figure \ref{fig:syn-vel}, this corrected Wittmann relation agrees well with the JPL ephemeris data and does not result in discontinuities. In Figure \ref{fig:syn-res} we compare residual errors for different methods. Method by Wittmann shows a maximum error of $\sim 0.008^{\circ}$ day$^{-1}$, while others are about ten times smaller. The residual error in method by \inlinecite{Rosa1995} is due to the usage of approximate formulae for the calculation of Earth's orbital elements, while one-month variations in the corrected Wittmann method are most probably due to the effect of the Moon which was neglected in Equation (\ref{e:omega}) for the calculation of the Earth's orbital angular velocity. When full JPL ephemeris are used and $\omega_{Earth}$ is calculated from Earth's linear orbital velocity or from the daily change of $\lambda_0$, these variations disappear.

Equation (\ref{e:this}) can also be derived analytically. Since instantaneous synodic-sidereal correction is simply given by the temporal derivative of the Earth heliographic longitude, Equation (\ref{e:corr}), there are two ways to calculate it. First is by estimating $L_0$ at two consecutive times, which leads to Equation (\ref{e:rosa}). The second one is by differentiating $L_0$. We start with the relation for $L_0$ ({\it e.g.}, \opencite{Meeus1991}; \opencite{Rosa1995}):
\begin{equation} \label{e:a1}
  \tan (L_0+W)=\tan (\lambda_0-\Omega)\cos i,
\end{equation}
where $W$ is the angular distance of the Carrington meridian from the ascending node. Temporal derivative of $L_0$ is:
\begin{equation} \label{e:a2}
  \frac{dL_0}{dt}+\frac{dW}{dt}=\frac{d}{dt}\left \{ \arctan[\tan (\lambda_0-\Omega)\cos i]\right \} .
\end{equation}
By using Equation (\ref{e:corr}) and $\omega_C=dW/dt$ instead of the left-hand side term, and calculating the derivative of the right-hand side term (neglecting $d\Omega/dt \approx 50$ arcsec/year), we get:
\begin{equation} \label{e:a3}
  \omega_{sid}-\omega_{syn}=\frac{\cos i}{[1+\tan^2(\lambda_0-\Omega)\cos^2 i]\cos^2(\lambda_0-\Omega)}\frac{d\lambda_0}{dt}.
\end{equation}
The last term on the right-hand side, $d\lambda_0/dt$, represents Earth's angular orbital velocity, $\omega_{Earth}$. Sorting out the denominator gives: 
\begin{equation} \label{e:a4}
  \omega_{sid}-\omega_{syn}=\omega_{Earth}\frac{\cos i}{\cos^2(\lambda_0-\Omega)+\sin^2(\lambda_0-\Omega)\cos^2 i},
\end{equation}
while converting sine to cosine and rearranging the denominator leads back to Equation (\ref{e:this}).

\section{Conclusion} 
      \label{S-Conclusion} 
Several methods for the calculation of the sidereal solar rotation velocity from the observed synodic velocity were investigated. Using state of the art ephemeris data, we found that the method by \inlinecite{Rosa1995} (Equation (\ref{e:rosa})) agrees well with JPL ephemeris calculations, while the method by \inlinecite{Wittmann1996} (Equation (\ref{e:wittmann})) shows some discrepancy and was later corrected (Equation (\ref{e:this})). Although the error in using incorrect Equation (\ref{e:wittmann}) is only of the order $0.01^{\circ}$ day$^{-1}$, this difference can be important in studies of all phenomena related to solar rotation and their time variation. This is especially important when the measured values are of this order of magnitude. 

The precision of the synodic-sidereal correction depends on the precision of $L_0$ and, consequently, $\lambda_0$, and numerical evaluation of their derivatives. Ephemeris systems and almanacs usually list $L_0$ with precision of $0.01^{\circ}$. Since synodic velocity is a temporal derivative of $L_0$, the precision of synodic-sidereal correction should only be affected by time variable terms which are usually given with much higher precision. The ultimate precision on the synodic-sidereal correction is also related to the uncertainties of the Carrington elements ($i$ and $\Omega$) ({\it e.g.}, \opencite{Balthasar1986}; \opencite{Balthasar1987}; \opencite{Beck2005}). Thus, by using Equations (\ref{e:rosa}) and (\ref{e:this}) with best available ephemeris, maximum precision is reached. If approximate formulae are used as given in \inlinecite{Rosa1995} and Equation (\ref{e:omega}) then the precision of synodic-sidereal correction is of the order of $0.001^{\circ}$ day$^{-1}$.

We advise to use any of the Equations (\ref{e:rosa}) \cite{Rosa1995} or (\ref{e:this}) (Wittmann corrected). An even better method would be transforming the observed central meridian distances to a coordinate system which is defined and fixed with solar rotation axis and then directly calculating the sidereal velocities without the need of synodic to sidereal transformation. Finally, we note that the original Wittmann equation (Equation (\ref{e:wittmann})) would give substantially larger errors for greater inclinations than in the Sun-Earth case. 
      


\begin{acks}
The authors used {\sc matplotlib} environment  \cite{Hunter2007} in {\sc python} to produce most of the plots in the paper. This work was partly funded from the European Commission FP7 projects eHEROES (284461, 2012-2015) and SOLARNET. SOLARNET is an Integrated Infrastructure Initiative (I3) supported by the European Commission's FP7 Capacities Program for the period April 2013 -- March 2017 under the Grant Agreement number 312495. The authors also thank the anonymous referee for insightful comments.
\end{acks}

   
\bibliographystyle{spr-mp-sola-cnd} 

\bibliography{synsid}  

\IfFileExists{\jobname.bbl}{} {\typeout{}
\typeout{****************************************************}
\typeout{****************************************************}
\typeout{** Please run "bibtex \jobname" to obtain} \typeout{**
the bibliography and then re-run LaTeX} \typeout{** twice to fix
the references !}
\typeout{****************************************************}
\typeout{****************************************************}
\typeout{}}

\end{article} 

\end{document}